\documentclass[journal=jacsat,manuscript=article]{achemso}

\usepackage{chemformula} % Formula subscripts using \ch{}

\usepackage{mhchem}
\usepackage{fancyhdr}
\usepackage{extramarks} 
\usepackage{amsmath}
\usepackage{amsthm}
\usepackage{amsfonts}
\usepackage{siunitx}
\usepackage{tikz}
\usepackage[plain]{algorithm}
\usepackage{algpseudocode}
\usepackage{multirow}
\usepackage{booktabs}
\usepackage{palatino}
\usepackage{graphicx}
\usepackage{subfigure}
\usepackage[colorlinks,linkcolor=black,anchorcolor=black,citecolor=black,urlcolor=blue]{hyperref}
\usepackage{amsmath,bm}
\usepackage{booktabs}
\usepackage{mathtools}
\usepackage{amssymb}
\usepackage{caption}
\usepackage{capt-of}
\usepackage{mciteplus}
\usepackage{mathrsfs}
\usepackage{makecell}
\usepackage[title,titletoc,toc]{appendix}
\usepackage{xr}
\usepackage{parskip}
\usepackage{soul}
\usepackage{textcomp}
\usepackage[colaction]{multicol}
\usepackage[switch]{lineno}
\usepackage{lipsum}
\usepackage{etoolbox}
\usepackage{longtable}
\usepackage{array}
\usepackage{tablefootnote}
\usepackage{graphicx}
\usepackage{color, colortbl}
\usepackage{hhline}
\usepackage[T1]{fontenc}

\newcommand{\crpartial}{\textup{\rmfamily\dh}}

\definecolor{Lightblue}{rgb}{0.867,0.914,0.961}
\definecolor{Lightgreen}{rgb}{0.883,0.934,0.848}

\newcolumntype{C}[1]{>{\centering\arraybackslash}p{#1}}
\usetikzlibrary{automata,positioning}
\usetikzlibrary{automata,positioning}

\author{Hongsong Feng}
\affiliation{Department of Mathematics, Michigan State University, MI 48824, USA.}

\author{Guowei Wei}
\affiliation
{Department of Mathematics, Michigan State University, MI 48824, USA.}
\alsoaffiliation
{Department of Electrical and Computer Engineering,
	Michigan State University, MI 48824, USA.}
\alsoaffiliation
{Department of Biochemistry and Molecular Biology,
	Michigan State University, MI 48824, USA.}
\email{weig@msu.edu}

\title {Virtual screening of DrugBank database for hERG blockers
  	using topological Laplacian-assisted AI models}

\abbreviations{IR,NMR,UV}

\begin{document}


\begin{abstract}
The human {\it ether-a-go-go} (hERG) potassium channel (K$_\text{v}11.1$)  plays a critical role in mediating cardiac action potential. The blockade of this ion channel can potentially lead fatal disorder and/or long QT syndrome. Many drugs have been withdrawn because of their serious hERG-cardiotoxicity. It is crucial to assess the hERG blockade activity in the early stage of drug discovery. We are particularly interested in the hERG-cardiotoxicity of compounds collected in the DrugBank database considering that many DrugBank compounds have been approved for therapeutic treatments or have high potential to become drugs. Machine learning-based in silico tools offer a rapid and economical platform to virtually screen DrugBank compounds. We design accurate and robust classifiers for blockers/non-blockers and then build regressors to quantitatively analyze the binding potency of the DrugBank compounds on the hERG channel.  Molecular sequences are embedded with two natural language processing (NPL) methods, namely, autoencoder and transformer. Complementary three-dimensional (3D) molecular structures are embedded with two advanced mathematical approaches, i.e.,  topological Laplacians and algebraic graphs. With our state-of-the-art tools, we reveal that 227 out of the 8641 DrugBank compounds are potential hERG blockers, suggesting serious drug safety problems. Our predictions provide guidance for the further experimental interrogation of DrugBank compounds'  hERG-cardiotoxicity . 
\end{abstract}

\section{Introduction}
The human  {\it ether-a-go-go}  related gene (hERG) encodes the protein (K$_\text{v}11.1$), the alpha subunit of a potassium ion channel, which is critical in the mediation of the cardiac action potential and the coordination of heartbeat. The blockade of this potassium channel is associated with prolongation of the QT interval (long QT syndrome, LQTS), eventually leading to fatal arrhythmia, namely Torsade de Pointes (TdP) \cite{shan2022review}. The hERG potassium channel can be blocked by many structurally and therapeutically diverse small compounds. As such, drug-induced cardiotoxicity was the main reason for drug withdrawals due to their unexpected hERG blockade activity.  For the concern of drug cardiac safety, it is desirable to have an early assessment of hERG liability in the process of drug design and development. In the early 2000s, the U.S. Food and Drug Administration (FDA) recommended the hERG side effect evaluation of drug candidates in their revised regulatory guidelines \cite{food2005international}. 

A variety of in vitro tests are available on the markets, such  as patch clamp techniques \cite{meyer2004micro}, radioligand binding \cite{finlayson20013h}, cell-based fluorescence \cite{dorn2005evaluation} and $^{86}$Rb flux assays \cite{cheng2002high}. Reliable hERG-affinity results of compounds could be obtained from those experimental methods. However, such assay tests tend to be expensive, time-consuming, and labor-intensive, rendering them unsuitable for screening a large collection of drug candidates in drug discovery. The development of in silico hERG models provides accurate computational predictions, offering fast and cheap approaches for virtual high-throughput screening. At present,  structure-based and ligand-based approaches are mostly used for developing in silico models for hERG blockade assessment. Structure-based models rely on the  structure of the potassium channel. To date, the X-ray crystal structure of hERG is not yet available, but structure from electron microscopy was solved recently \cite{wang2017cryo}. Previous structure-based studies performed their investigations utilizing homology modeling of hERG channel based on the available (crystal) structure of other potassium channels \cite{doyle1998structure, jiang2002crystal, zhou2001chemistry,dempsey2014assessing,kalyaanamoorthy2018binding} in conjunction with docking simulations and free energy calculations. Recent microscopy structures prompted more mechanism  investigations of hERG-drug interaction \cite{furutani2022facilitation,emigh2019structural} and arrhythmogenicity predictions \cite{cortez2022predicting}. Molecular blocking activities on hERG positively were found to be related to charged nitrogen atoms with aromatic or hydrophobic groups \cite{farid2006new}, hydrogen-bond acceptor at the periphery \cite{perry2006drug} or high lipophilicity \cite{waring2007quantitative}. The development of structure-based models faces challenges from protein flexibility and poor scoring functions \cite{jia2008binding,li2013id}. Such models mostly investigate the hERG-drug interactions around the hERG central cavity involving several residues \cite{vandenberg2017towards,maly2022structural} responsible for drug binding. Multiple other binding sites also exist for hERG blockers with different structural features \cite{vandenberg2017towards}, leading to difficulty in developing reliable prediction models \cite{aronov2005predictive}. 

Many ligand-based models have been proposed over past few decades by utilizing quantitative structure-activity relationship studies, pharmacophore modeling, and machine learning algorithms. Machine learning models draw increased popularity for the hERG blockade predictions with their predictive ability advancement supported by the significant increases in bioactivity information about hERG inhibitors in public databases (e.g., ChEMBL, PubChem) and literature. Among recent machine learning models of predicting hERG blockade, in 2010, Doddareddy and co-workers applied linear discriminant analysis and support vector machine (SVM) methods to build classification models on a dataset of 2,644 compounds \cite{doddareddy2010prospective}. Extended connectivity fingerprints (ECFPs) and functional class fingerprints (FCFPs) were used to describe molecule compounds. In 2014, Liu \cite{liu2014novel} also developed a Bayesian classification model based on the Doddareddy et al.'s dataset with a series of molecular properties and extended-connectivity fingerprints (ECFP$_4$).
In 2016, Didzidapetris et al. employed gradient boosting machine to build their classification model on a large and chemically diverse hERG inhibition dataset of 6690 compounds. Simple physicochemical and topological parameters of molecules were used as descriptors for the machine learning models. In 2020, Wang et al.\cite{wang2020capsule} used a deep learning network algorithm, capsule network (CapsNets), for classification models. Their model further boosted the prediction accuracy for the test set of  Doddareddy et al. Many hERG inhibitor datasets were collected and machine learning models of pronounced predictive powers were proposed. Recently, some even larger hERG inhibitor datasets  \cite{ogura2019support,zhang2022hergspred} with rich chemical or structure diversity were compiled through the integration of multiple databases and literature. Zhang et al. \cite{zhang2022hergspred} collected a dataset of 12,850 compounds and built a consensus model using random forest, extreme gradient boosting tree algorithm, and deep neural network with two-dimensional (2D) fingerprints as molecular descriptors. Ogura et al. \cite{ogura2019support} and his coworkers integrated a dataset of more than 291,000 structurally diverse compounds derived from ChEMBL, GOSTAR, PubChem, and hERGCentral. Their classification model based on the support vector machine (SVM) algorithm achieved an accuracy of 0.984 for their test set. Such a large hERG inhibitor dataset with high molecular structure diversity broadens the predictability and scope of machine learning models.

The development of machine learning models makes it possible to have a high-throughput screening  in the early stage of drug discovery or in reevaluating the hERG-related cardiotoxicity of on-market drugs. DrugBank database (version 5.1.9) has a collection of 8641 compounds that are mostly either FDA-approved, experimental, or investigational drugs. It gains wide applications in in silico drug discovery, design, drug docking or screening \cite{wishart2008drugbank}. In particular, the repositioning potentials of DrugBank compounds were studied in various aspects including those for COVID-19 \cite{gao2020repositioning,beigel2020remdesivir}. Due to its important role in drug discovery, studies on the hERG side effect receive particular attention. The development of machine learning models with a wealth of available hERG inhibitor datasets facilitates the screening DrugBank database for hERG blockers. This motivates us to develop more robust and accurate models to analyze the blockade potential of DrugBank compounds.

The performance of machine learning models relies on molecular embeddings or descriptors. Many of the aforementioned machine learning models \cite{doddareddy2010prospective,wang2012admet,liu2014novel,zhang2022hergspred,ogura2019support} or other  hERG blockade prediction models in the literature employed  2D fingerprints as   molecular descriptors.  Two-dimensional embeddings encode interpretable physical properties in a bit string format while three-dimensional (3D) embeddings could preserve 3D molecular structural patterns and most importantly,  stereochemical information. Deep neural network (DNN)-based techniques have been utilized to exploit the molecular descriptors in embedded vector forms. Recently, sequence-to-sequence autoencoders with translation methodology were used to develop molecular embeddings \cite{winter2019learning}. The encoder network compresses comprehensive information of the SMILES string in latent vectors, which are then translated to another form of molecular representation by the decoder network. Besides, we recently developed a self-supervised learning (SSL) Transformer platform to extract useful physicochemical information from the SMILES of millions of compounds in various databases \cite{chen2021extracting}. The bidirectional encoder Transformer (BET) based on self-attention was used to achieve SSL. These natural language processing (NPL) techniques extract physicochemical  information by investigating molecular sequence patterns. However, the intricate structural complexity of small molecules that encode sufficient stereochemical information usually cannot be well embedded by sequence-based methods. Three-dimensional (3D) embeddings were developed to preserve enough 3D molecular structural patterns. Advanced 3D molecular embeddings were devised with topological Laplacians \cite{wang2020persistent}. The harmonic spectra  and non-harmonic spectra of topological Laplacians embed respectively topological persistence and geometric shape into graph invariants along a  series of filtration parameters. It showed the superior descriptive and predictive ability of molecular complexes in the infectivity prediction of the SARS-CoV-2 Omicron variants \cite{chen2022persistent}. Algebraic graph learning also provides powerful molecular descriptors by interpreting the molecular 3D structures in algebraic graphs \cite{nguyen2019agl}. 

In this work, we focused our attention on the hERG liability screening of DrugBank compounds by advanced machine learning models.  Two natural language processing (NPL)-based latent-space embeddings and two advanced mathematics-based 3D embeddings were integrated with gradient boosting decision tree and/or deep neural network algorithms. The two sequence-based embeddings were constructed  using Transformer \cite{chen2021extracting} and autoencoder \cite{winter2019learning} methods while the two 3D embeddings were generated with topological Laplacians \cite{wang2020persistent,chen2022persistent} and algebraic graphs  \cite{nguyen2019agl}. Their combined efforts in extracting molecular physical and stereochemical information facilitate enhanced descriptive and predictive power for virtual screening of small molecules. The binary classification model was first used to predict the potential hERG blockers from DrugBank compounds and then the regression model was employed to quantitatively analyze the binding affinity of these blockers. According to our classification model, 227 out of DrugBank compounds were predicted to be blockers, among which 92 are FDA-approved drugs and 135 are investigational drugs. A few of the 227 drugs have been withdrawn for various reasons including cardiotoxicity. Except for these withdrawn drugs, further cardiotoxic risk assessment can be carried out for these predicted blockers. Further experimental investigation can be carried out to test the QT prolongation effect and scrutinize their cardiotoxicity in humans. Compared to the literature methods, our models demonstrated the state-of-the art predictive power. Our work unveils potentially serious drug safety problems.

\begin{figure}[ht]
	\centering
	\includegraphics[width=0.57\linewidth]{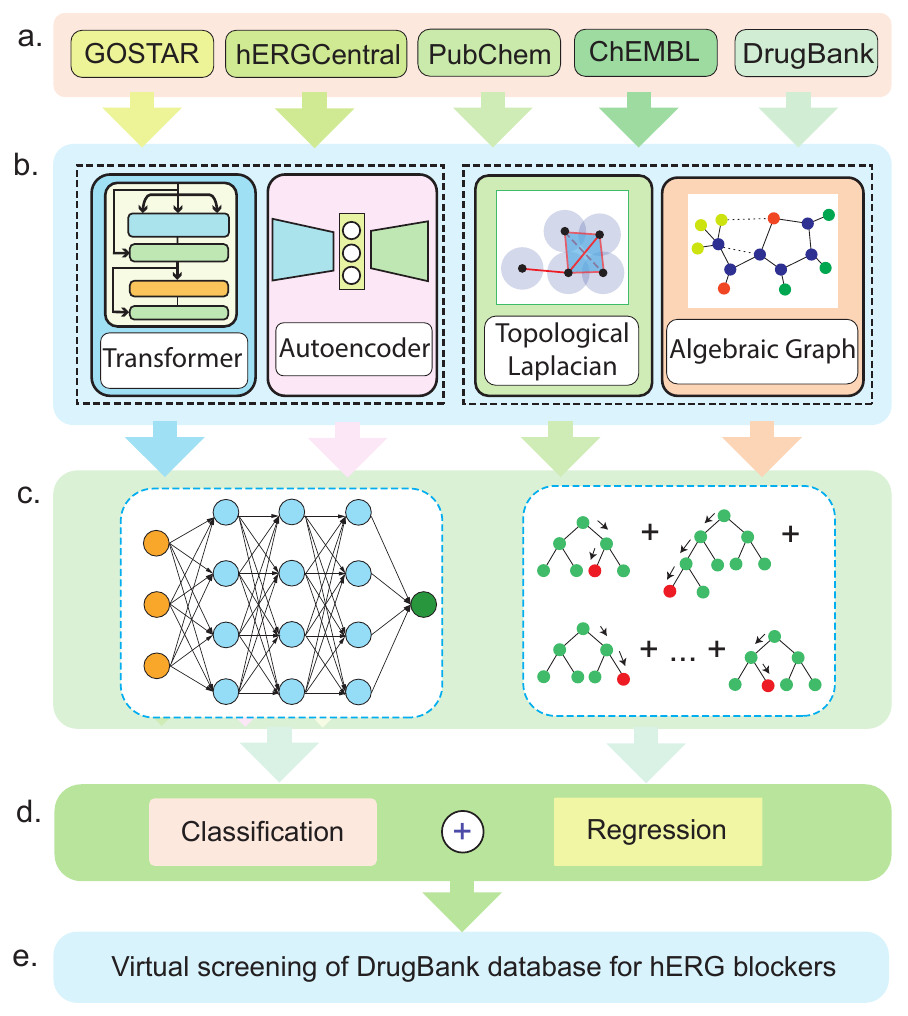} 
	\caption{  Illustration of our machine learning platform for screening the DrugBank database for hERG blockers. a. Datasets from GOSTAR, hERGcentral, PubChem, ChEMBL databases are used to build classification and regression models. Compounds in DrugBank database are  used  in our hERG blockade screening. b. Molecular embeddings for compounds from above databases are generated by transformer, autoencoder, and topological Laplacian, and algebraic graph algorithms. Natural language processing (NPL)-based Transformer and autoencoder offer sequence information. Advanced mathematics-based topological Laplacian and algebraic graph provide complementary 3D structural information. c. Gradient boosting decision tree (GBDT) and deep neural network (DNN) algorithms are integrated with the aforementioned molecular embeddings to build machine learning models. d. Classification and regression models are trained by using the aforementioned molecular embeddings and machine learning models. e.  Virtual screening of the DrugBank database is carried out for hERG blockers using the trained classification and regression models.}  
	\label{Fig:workflow}
\end{figure}

\section{Results}

We are interested in accurate and reliable screening of the hERG blockade of the compounds deposited in the DrugBank database (version 5.1.9). To this end, we investigated the performance of our machine-learning classification model on seven different datasets. Among them, the training dataset from Ogura et al. \cite{ogura2019support} consists of more than 200,000 of compounds, covering molecules of large structural and chemical diversities. This is a classification dataset with only ``yes'' or ``no'' labels. Machine learning models based on such a large training dataset can deliver high valuable information in practical applications. % Hence our machine learning model built on this dataset was employed to predict the hERG blockade activity of drugs in the DrugBank database. 
Our classification model  achieves the accuracy of 0.981 for the test set from Ogura et al. \cite{ogura2019support}, indicating the high reliability of our models. The threshold of 10 $\rm \mu$M was used to discriminate compounds into hERG blockers or non-blockers in Ref. \cite{ogura2019support}. This threshold was adopted in our discussion to determine whether DrugBank compounds are  hERG blockers or not. In addition to the qualitative analysis using our classification model, we also build a quantitative assessment of the binding affinity (BA) of the DrugBank compounds. We collected an hERG inhibitor dataset composed of 6298 compounds from the ChEMBL database and the literature. The experimental binding affinities of those inhibitors were quantified by either $K_i$ or $\rm IC_{50}$. As suggested in Ref. \cite{kalliokoski2013comparability}, the $\rm IC_{50}$ values can be approximately converted to $K_i$ values by the formula $ K_i=\rm IC_{50}/2$. We determined the labels of our collected dataset of hERG inhibitors with binding affinity calculated by  using $1.3633\times \log_{10}K_i.$ With the collected dataset, we used three types of molecular descriptors, namely those from transformer, autoencoder, and topological Laplacian, in conjunction with the gradient boosting decision tree algorithm to build three regression models. The consensus results given by these three models were used  as the final results. Besides, each model was run twenty times with different random seeds such that the average of twenty results was regarded as the prediction of each model. The performance of our regression model was validated by five-fold cross validation, which showed a Pearson correlation coefficient of 0.77 and root-mean-square-error (RMSE) of 0.796 kcal/mol.

The DrugBank database curated a total of 8641 drugs, among which 1782 drugs were approved by FDA and other 6859 are experimental or off-market drugs. To investigate the hERG liability of the DrugBank compounds, we first used our classification model to determine the potential hERG blockers and then employed our regression model to measure their binding affinities. According to our classification model, a total of 227 drugs were classified as potential blockers, among them,  92 are FDA-approved drugs and other 135 are experimental or off-market drugs. Based on our regression model, we report the top 20 FDA-approved drugs with the ranked binding affinity values in the Table \ref{tab:drugbank-approved}. Since some of FDA-approved or experimental drugs were withdrawn from the market, we are also interested the hERG liability of the withdrawn drugs. Table \ref{tab:drugbank-approved} also includes four additional drugs that were approved by FDA but withdrawn later. The predicted binding affinities (kcal/mol) are listed in the table while the predicted $\rm IC_{50}$ were derived from the predicted binding affinities with the formula $\rm IC_{50}=10^{\frac{2 X}{1.3633}}\times 10^{6}$ ($\mu$M) where $X$ is the binding free energy. A complete list of binding affinities for the 92 approved drugs can be found in the Supporting information. The threshold of 10$\mu$M was used for classifying hERG blockers/non-blockers in our investigation. Our predicted $\rm IC_{50}$ for 85 out of the 92 potential blockers are less than 10 $\mu$M, which shows the prediction consistency between our classification and regression models. The corresponding binding affinity threshold is equal to -7.23 kcal/mol based on the aforementioned conversion formula.

\subsection{Potential hERG blockers}

We next present a brief discussion about the top 20 potential blockers and four FDA-approved but withdrawn drugs. The top 20 blockers were predicted to have potent binding affinity compared to the threshold BA of -7.23 kcal/mol. Ibutilide was predicted to be the most potent blocker. Ibutilide is a class III antiarrhythmic medication used to correct atrial fibrillation and atrial flutter to sinus rhythm. Studies have found that Ibutilide can lead to abnormal heart rhythms because of its ability to prolong the QT interval, potentially leading to Torsades de Pointes,   a type of very fast heart rhythm (tachycardia), which is closely related to its block on hERG channel \cite{perry2004structural}. Our prediction of Ibutilide is validated by the experimental study and clinical findings. The second blocker is Dofetilide, which is also a class III antiarrhythmic agent. Similar to   Ibutilide,   Dofetilide is also used to for cardioconversion from atrial fibrillation and atrial flutter to sinus rhythm. It can also cause side effect of Torsades de Pointes, and the dose has to be administrated and adjusted carefully in patients by monitoring the corrected QT. The third drug is Lidoflazine, which is a vasodilator used to treat angina pectoris. It can bind to the hERG channel with a high binding affinity \cite{katchman2006comparative}. The   fourth drug is Pimozide, which is an antipsychotic drug. It can cause a number of side effects including prolongation of the QT interval and binds to hERG with $K_i$=18nM \cite{goodman1996goodman}. Ziprasidone is an atypical antipsychotic used for the treatment of  schizophrenia and bipolar disorder. It is thought to possess lower side effects than other antipsychotic medications. It raised concerned about long QT syndrome considering its binding affinity of $K_i$ equal to 169 nm \cite{kongsamut2002comparison}. Our predicted $\rm IC_{50}$ for Ziprasidone is 0.15 $\mu$M with the derived $K_i$ value of 75 nM, which is at the same magnitude as the experimental $K_i$ value. Domperidone is a dopamine antagonist drug that is used to treat nausea, vomiting and gastrointestinal problems. It can cause QT prolongation and is associated with the increased risk of sudden cardiac death \cite{leelakanok2016domperidone}. With our model, Domperidone was predicted to block hERG potassium channels and possess a moderately high binding affinity of IC$_{50}$=0.17$\mu$M. Ebastine is a H1 receptor antagonist used in the treatment of allergic rhinitis and chronic idiopathic urticaria. It gives an overall favorable safety profile with no QT prolongation \cite{gillen2001effects}. Based on our prediction, Ebastine is anticipated to bind to hERG channel. Amsacrine is an antineoplastic agent and exert its effects by binding to DNA. Our prediction indicates its blockade to hERG channel, agreeing with the report that it can cause inhibition of hERG currents \cite{thomas2004inhibition}. The next one is Benzethonium, often used as disinfection and surface treatment in hospitals and industries. It was reported to induce the long QT syndrome and exhibit hERG inhibition \cite{long2013mechanism}. Dronedarone is also a medication for cardiac arrhythmias. Torsade de Pointes tachycardia was observed in dronedarone therapy \cite{huemer2015torsade}.

The eleventh drug is Atracurium. Atracurium is a neuromuscular blocking agent and is widely used in anesthesia. Its common side effects include flushing of the skin and low blood pressure, which is mainly explained by the histamine release. Our prediction indicates its risk in the blockade of the hERG potassium channel. The next agent is Mivacurium, which is in the same family as Atracurium and is also used in anesthesia. The thirteenth drug is Ubiquinol, which can be used to protect cells from oxidative damage and sustain the effects of vitamin E. It can be synthesized by the human body. Fluspirilene is an antipsychotic drug used in the treatment of schizophrenia. Fluspirilene was found to be a Na$^+$ channel inhibitor and can induce arrhythmia in drug treatment \cite{sampurna2019cardiac}. Bazedoxifene is a medication used in the prevention of postmenopausal osteoporosis and is investigated for the possible treatment of cancer. Sunitinib is a medication used to treat cancer. It was approved for the treatment of two different cancers including renal cell carcinoma and imatinib-resistant gastrointestinal stromal tumor. It is not associated with serious adverse effects in its application. Umeclidinium was approved for the maintenance treatment of chronic obstructive pulmonary disease. There was a controversy on its relation to fatal and nonfatal cardiovascular toxicity \cite{tashkin2015safety}. Oxatomide is an antihistamine drug used to treat and prevent allergic symptoms. Trials studies were also carried out for the treatment of Duchenne muscular dystrophy. Afamelanotide is a subcutaneous implant used to protect people with erythropoietic protoporphyria from phototoxicity. No report on its cardiotoxicity was found. Iloperidone is in the same family of Ziprasidone as an atypical antipsychotic for the treatment of schizophrenia. Like Ziprasidone, Iloperidone was reported to be a potent hERG blocker and delays cardiac ventricular repolarization \cite{vigneault2012iloperidone}. It is in line with our prediction that Iloperidone can be a blockers with high binding affinity. The six top-ranked drugs among these 20 FDA-approved drugs have reported high binding affinities to the hERG channel or QT prolongation effect. These six drugs are dedicated to either cardiac diseases or psychotic disorders. The top ten blockers have reported high binding affinities or are related to long QT syndrome. Some of the eleventh to twentieth blockers were reported to be associated with cardiotoxicity. Our prediction reveals that  other predicted blockers need to be tested on their cardiotoxicity.

Some FDA-approved drugs were withdrawn for a variety of reasons. In our prediction, four such drugs were predicted to be hERG blockers with moderate binding affinities. Sertindole is an antipsychotic medication and was used in the treatment of schizophrenia. It was withdrawn owing to the increased risk of sudden death from QTc prolongation in the USA in 1998 and is a high affinity antagonist of hERG potassium channel. Terfenadine was used in the treatment of allergic conditions. It was withdrawn because of cardiac arrhythmia caused by QT interval prolongation. Indoramin is an antiadrenergic drug with direct myocardial depression action, resulting in no reflex tachycardia. Cisapride is used to treat gastroesophageal reflux disease and was withdrawn due to serious cardiac side-effects. The four drugs were all withdrawn due to their cardiotoxicity issues. Therefore, the predictions of our models are in line with the withdrawal reasons for the four drugs. In particular, several of the listed 20 drugs are antipsychotics. Clinical statistics have revealed a close correlation between the use of antipsychotics and the increased risk of cardiac issues or deaths. This suggests the importance of cardiotoxicity prediction in drug design of antipsychotics.

In addition to the aforementioned 92 FDA-approved drugs with predicted hERG-blocking effects, our predictions also gave a total of 135 investigational or off-market drugs that can potentially block the hERG channel. Twenty top-ranking investigational drugs and three withdrawn drugs are shown in Table \ref{tab:drugbank-investigational}. Among the 135 investigational drugs, 133 drugs were predicted to have binding affinity values less than -7.23 kcal/mol. The most potent drug is DB12517. It has been investigated for the treatment of erectile dysfunction. Our prediction suggested its high risk of causing heart problems. The withdrawn drugs include ifenprodil, mibefradil, and eprazinone.  Ifenprodil has been investigated for preventing tinnitus after acoustic trauma. Especially,  it is now in phase III clinical trials for the treatment of SARS-CoV-2 infection. A complete list of the predicted binding affinity values for the 135 drugs can be found in the Supporting information. Our prediction offers a warning for possibly causing heart problems when applying the investigational drugs in treatment. 

\begin{table}
	\centering
	\small
	\begin{tabular}{c| c c c c c c}		
		\toprule
		&\makecell[c]{Drugbank\\ID} & \makecell[c]{generic \\name} &\makecell[c]{ predicted \\BA}	& \makecell[c]{predicted \\$\rm IC_{50}$} &\makecell[c]{regression \\reliability} &\makecell[c]{classif.\\reliability} \\
		\hline 
		1 &DB00308  & Ibutilide &-11.06  &  0.02 & 1.0&0.81\\
		2 &	DB00204  &Dofetilide & -11.04 & 0.02 & 1.0 &1.0\\
		3 &DB13766 & Lidoflazine &  -11.01 &  0.02 & 1.0& 0.84\\
		
		4 &DB01100 & Pimozide &-11.06   & 0.08 &1.0 &0.87\\
		
		5 &DB00246  & Ziprasidone & -9.72 & 0.15& 1.0 &1.0\\
		
		6 &DB01184  & Domperidone & -9.64 & 0.17& 1.0 &0.87\\
		
		7 &DB11742 & Ebastine & -9.62& 0.18& 1.0&0.92\\
		
		8 &DB00276  & Amsacrine & -9.5& 0.22& 1.0&1.0\\
		
		9 &DB11125  & Benzethonium & -9.35 & 0.28& 0.94&1.0\\
		
		10 &DB04855  & Dronedarone & -9.27& 0.32& 1.0&1.0\\
		
		11 &DB00732  & Atracurium & -9.21  & 0.35& 0.79&0.74\\
		
		12 &DB01226  &Mivacurium & -9.14 & 0.39& 0.78&0.78\\
		
		13 &DB11340 & Ubiquinol & -9.12  & 0.39& 0.61&0.7\\
		
		14 &DB04842 & Fluspirilene &  -9.09& 0.43& 0.87&1.0\\
		
		15 &DB06401  & Bazedoxifene & -8.98 & 0.52& 0.77&0.77\\
		
		16 &DB01268  & Sunitinib & -8.96 & 0.53& 1.0&1.0\\
		
		17 &DB09076 & Umeclidinium & -8.95& 0.54& 0.85&0.85\\
		
		18 &DB12877  & Oxatomide& -8.85& 0.65& 0.88&0.83\\
		
		19 &DB04931  & Afamelanotide& -8.82& 0.68& 0.65&0.68\\
		
		20 &DB04946  & Iloperidone& -8.77 & 0.74& 0.86&0.87\\
		\cmidrule(l{1em}r{1em}){1-7}
		
		1 &DB06144  & Sertindole &  -11.1 & 0.01 & 1.0&1.0\\
		
		2 &DB00342  & Terfenadine & -8.87 &  0.62& 1.0&0.93\\
		
		3 &DB08950  & Indoramin & -8.55& 1.07 &0.8 &1.0\\
		
		4 &DB00604 & Cisapride & -8.26& 1.75& 0.94 &0.99\\
		
		\bottomrule
	\end{tabular}
	\caption{Summary of the predicted top 20 potential hERG blockers from FDA-approved drugs and four additional approved drugs that have been withdrawn from the market. The predicted binding affinities (unit: kcal/mol) and $\rm IC_{50}$ ($\mu$M) values by our regression model are given. Additionally, the reliability scores of the regression and classification prediction are provided.}
	\label{tab:drugbank-approved}
\end{table}

\subsection{Prediction reliability  }

In the last subsection, forty FDA-approved or investigational drugs and some withdrawn drugs were presented with their predicted binding affinity to hERG. These DrugBank compounds were classified into the category of hERG blockers according to our classification models. The classification and regression predictions showed high consistency with the threshold of IC$_{50}$. To further analyze the reliability of our predictions, we study the reliability scores of each prediction from our classification or regression models as shown in the last two columns of Table \ref{tab:drugbank-approved} and \ref{tab:drugbank-investigational}. The reliability scores were calculated based on compound similarities with training datasets. The details of similarity calculation can be found in the Supporting information. 

The similarity scores can be equal to 1 in our calculations. This is because some of the DrugBank compounds are in our classification or regression datasets. The classification category or the binding affinity values can be well predicted in this scenario. For a given molecule, the higher the reliability scores, the more reliable our predictions are. The top 10 drugs in Table \ref{tab:drugbank-approved} showed reliability scores equal to or close to 1 for the regression predictions while the corresponding classification reliability scores are also high. As discussed before, the top 10 drugs except ebastine were reported to be associated with QT prolongation effect. This is can be well explained considering that they inhibit the hERG potassium channel. Ebastine is an exception. No clinically relevant changes in QTc interval were observed when ebastine was exercised at a higher dose than recommended therapeutic level  \cite{gillen2001effects}. It does not conflict our prediction from our regression and classification models as ebastine was reported to suppress the mammalian hERG potassium channel. The high-reliability scores for both regression and classification predictions reflect the high likelihood that the ten drugs are hERG potassium blockers. It is known that drugs that induce fatal arrhythmia or TdP mostly inhibit the hERG potassium and prolong QT interval. But the converse is not always true \cite{finlayson2004acquired}. It deserves further scrutiny on potential QT interval prolongation even if a compound was predicted as a potent hERG channel inhibitor. The number eleven to twenty drugs were predicted to be potent hERG blockers with moderately high-reliability scores. Among these ten drugs, fluspirilene, umeclidinium, and iloperidone have been reported to be associated with cardiac issues and were found to inhibit the human hERG potassium channel.  

No QT prolongation effect was reported for the seven other drugs. Sunitinib was found to be a potent inhibitor in both experiments and our predictions. The regression or classification reliability scores for atracurium, mivacurium, bazedoxifene, and oxatomide are close to or greater than 0.8. These four drugs have high chances to be potent inhibitors of the hERG potassium channel. In addition, the reliability scores for ubiquinol and afamelanotide are relatively low but still have classification reliability close to 0.7. The predictions indicate that these seven drugs deserve further experimental investigations on their binding affinity to the hERG potassium channel or their QT interval prolongation effect. The reliability scores for the four withdrawn drugs are high in both the regression and classification predictions. They were withdrawn due to cardiotoxicity issues, which agrees with the predicted high binding affinities.
However, the similarity scores for Ubiquinol and Afamelanotide are relatively low. Further investigations are needed to determine their hERG side effects. 

The prediction reliability scores were also provided for 20 investigational drugs in Table \ref{tab:drugbank-investigational}. For the investigational drugs with high-reliability scores, we may anticipate that they are highly likely to be hERG blockers and potentially cause cardiac issues.  More results on all the 227 predicted hERG blockers were provided in the Supporting information. The reliability scores give us a basic understanding of the blockade potentials. Experimental studies are required to further measure the hERG blockade and QT prolongation effect. 

Relatively low similarities with our training sets were found  for Octylphenoxy polyethoxyethanol, 
23,27,31-Octamethyldotriaconta-2,6,10, and Dopastatin. Their possible hERG side effects need to be further studied with other means.

\begin{table}
	\centering
	\small
	\begin{tabular}{c |c c c c c c}		
		\toprule
		&\makecell[c]{Drugbank\\ID} & \makecell[c]{generic \\name} &\makecell[c]{ predicted \\BA}	& \makecell[c]{predicted \\$\rm IC_{50}$} &\makecell[c]{regression \\reliability} &\makecell[c]{classif.\\reliability} \\
		\hline
		1 &DB12517 & PF-00446687 & -10.97 & 0.02 &0.99&0.97\\
		2 &DB07405 & \makecell[c]{1-(6-CYANO-3-PYRIDYLC\\ARBONYL)-5',8'-D\\IFLUOROSPIRO[PI\\PERIDINE-4,2'(\\1'H)-QUINAZOLINE]-\\4'-AMINE}& -9.92 & 0.11 &1.0 &1.0\\
		3 &DB04682 &\makecell[c] {Octylphenoxy \\polyethoxyethanol} & -9.51&0.21 & 0.63&0.71\\
		4 &DB03232 & \makecell[c]{2-[(2e,6e,10e,14e,18e,22e,26e)-\\3,7,11,15,19,23,27,31-\\Octamethyldotriaconta-\\2,6,10,14,18,22,26,\\30-Octaenyl]Phenol} & -9.51 & 0.21 &0.66 &0.86\\
		5 &DB05695 & NPS-2143 & -9.29 & 0.31 &1.0 &0.99\\
		6 &DB13791 & Penfluridol &- 9.12 & 0.41 & 0.84&0.84\\
		7&DB05414 & Pipendoxifene & -8.94 & 0.56 &0.76 &0.79\\
		8&DB16144 & Dopastatin & -8.88 &  0.61 & 0.63& 0.64\\
		9&DB13511 & Clebopride & -8.85& 0.65 &1.0 &1.0\\
		10&DB12869 & Eliprodil & -8.84 & 0.66 &0.83  &0.84\\
		11&DB04615 & (S)-tacrine(10)-hupyridone & -8.8&  0.7 &0.76 &0.78\\
		12&DB02615 & Compound 19 & -8.77& 0.74 &0.81&0.83\\
		13&DB08622 & \makecell[c]{4-(4-CHLORO-PHENYL)-1\\-\{3-[2-(4-FLUORO-PHENYL)-[\\1,3]DITHIOLAN-2-YL]-\\PROPYL\}-PIPERIDIN-4-OL}& -8.76& 0.75 &0.81 &0.81\\
		14&DB04614 & (R)-tacrine(10)-hupyridone & -8.75& 0.76 & 0.76&0.78\\
		15&DB06311 & Darapladib & -8.74& 0.77 &0.77 &0.77\\
		16&DB04471 &\makecell[c]{ 2-Phenyl-1-[4-(2-Piperidin\\-1-Yl-Ethoxy)-Phenyl]-\\1,2,3,4-Tetrahydro-\\Isoquinolin-6-Ol} & -8.73& 0.79 &0.8 &0.8\\
		17&DB02715 & Compound 18 & -8.7& 0.83 & 0.81 &0.83\\
		18&DB13554 & Moperone & -8.65& 0.9 &0.84 &0.88\\
		19&DB05137 & Lobeline & -8.65 & 0.9 &0.81 &0.94\\
		20&DB08009 & SU-11652 & -8.63 & 0.94 &0.99 &0.99\\
		\cmidrule(l{1em}r{1em}){1-7}
		1&DB08954 & Ifenprodil & -9.91&0.11&1.0&0.81\\
		2&DB01388 & Mibefradil & -8.77& 0.74 &1.0&0.97\\
		3&DB08990 & Eprazinone &-8.43 &1.32 &0.78 &0.82\\
		\bottomrule
	\end{tabular}
	\caption{Summary of the predicted top 20 hERG blockers from investigational or experimental drugs and three investigational or experimental drugs that have been withdrawn from the market. The predicted binding affinities (unit: kcal/mol) and $\rm IC_{50}$ ($\mu$M) values by our regression model are given. Besides, the reliability scores of the regression and classification prediction are provided.}
	\label{tab:drugbank-investigational}
\end{table}

\section{Discussion}

\subsection{Model performance comparison  }

In this study, we mainly used three types of molecular embeddings to build machine learning models and the details about the three embeddings are presented in the materials and methods section. The three embeddings were combined with gradient boosting decision tree (GBDT) and deep neural network (DNN) algorithms to develop six classification models. The consensus results were determined by the simple average of the six sets of blocker probability scores from the six models. Given a training dataset, twenty different random seeds were used to build each of the six individual models twenty times. In the comparison  between our models with the state-of-art in the literature, the highest evaluation metrics values of the twenty consensus results were reported. Seven hERG blockade datasets with binary classification labels from the literature were used to investigate the performance of our models. The details of these datasets can be found in the material and method section. Additionally, five evaluation metrics are also described in the Supporting information. Table \ref{tab:classication-results} shows the comprehensive comparisons between our models and other methods in the literature in terms of the five evaluation metrics.

\begin{table}
	\centering
	\small
	\begin{tabular}{c c c c c c c }		
		\toprule
		\multirow{2}{*}{\bf{Dataset}} & \multirow{2}{*}{\bf{Models}}	&\multicolumn{5}{c}{\bf{Metrics}}\\\cline{3-7}
		&    & {\bf AUC } &  {\bf  Accuracy} &  { \bf MCC}   & {\bf Sensitivity}  &{ \bf Specificity}   \\ 
		\hline
		\multirow{3}*{Braga et al.'s \cite{braga2015pred}} & Braga \cite{braga2015pred} &	NA& 0.740 &	NA	&	NA 	 &   NA   \\
		&X. Zhang et al.\cite{zhang2022hergspred}  & 0.891&  0.814 &0.627  &0.871 &0.766  \\
		&Our study  & {\bf0.892} & {\bf 0.814 }&  {\bf0.627}&      0.800&	0.827	 				 \\
		\hline
		
		\multirow{3}*{C. Zhang et al. \cite{zhang2016silico}} & C. Zhang \cite{zhang2016silico} & 0.798 & 0.847& 0.436 &0.989 &0.277  \\
		&X. Zhang \cite{zhang2022hergspred}   & 0.803 &  0.856 	&    0.445& 0.973 	& 	0.383 	 	 \\
		&Our study  & {\bf0.836} &{\bf 0.864 }&	 {\bf 0.518	}&	0.974	&	0.426	 \\
		\hline
		
		\multirow{3}*{Li et al. \cite{li2017modeling}} & Li \cite{li2017modeling} & 0.881 &  0.842  & 0.573 & 0.626 &  0.918  \\
		&X. Zhang \cite{zhang2022hergspred}  &0.873   &0.867 &   0.584 &	 0.564 	 &	0.950 	  	 \\
		&Our study  &{\bf 0.917} & {\bf0.885}&  {\bf 0.629}	 &0.573	&	0.970			 \\
		\hline
		
		\multirow{3}*{Cai et al. \cite{cai2019deep}} & Cai \cite{cai2019deep} & 0.967 &  0.925 & 0.579  &  0.926 &  0.914 \\
		&X. Zhang \cite{zhang2022hergspred}   &0.995 &  0.983 & 0.920  &0.986 & 0.966  \\
		&Our study & {\bf1.000} &  {\bf0.998}&  {\bf0.990	}	&  0.998	& 	1.000	 	 \\
		\hline
		
		\multirow{3}*{Doddaredy et al.\cite{doddareddy2010prospective}} & Doddaredy\cite{doddareddy2010prospective} &NA & 0.880  &      NA   &	NA 	 &   NA          \\
		&Wang\cite{wang2020capsule}  & 0.940 & 0.918&    0.835   &      0.944    &	0.898 	  \\
		&Our study & {\bf0.975} & {\bf0.922	} &	 {\bf0.840}	& 0.917	&	0.925  \\
		\hline
		
		\multirow{3}*{Ogura et al. \cite{ogura2019support}} & Ogura \cite{ogura2019support} & 0.962 & {\bf0.984 } &    NA&        0.670 & 0.995       \\
		&Chavan \cite{Chavan2022herg}  & 0.950 & NA  & NA& NA & NA   \\
		&Our study  & {\bf 0.963 } & 0.981 &	   0.662	 &     0.512	&0.997 \\
		\hline
		
		\multirow{2}*{X. Zhang et al. \cite{zhang2022hergspred}} & X. Zhang \cite{zhang2022hergspred}& NA & 0.839 & NA  & NA& NA \\
		&Our study  & 0.915 & {\bf0.842}& 0.683  & 0.842 & 0.842   \\
		\bottomrule
	\end{tabular}
	\caption{The prediction performance comparisons between our classification models and literature work on various hERG blockade datasets. The Braga et al.'s  \cite{braga2015pred} dataset is for five-fold cross validation task while all others are for prediction on test sets. NA means the predicted results were not reported from the corresponding literature publications.}
	\label{tab:classication-results}
\end{table}

\begin{figure}[ht]
	\centering
	\includegraphics[width=0.85\linewidth]{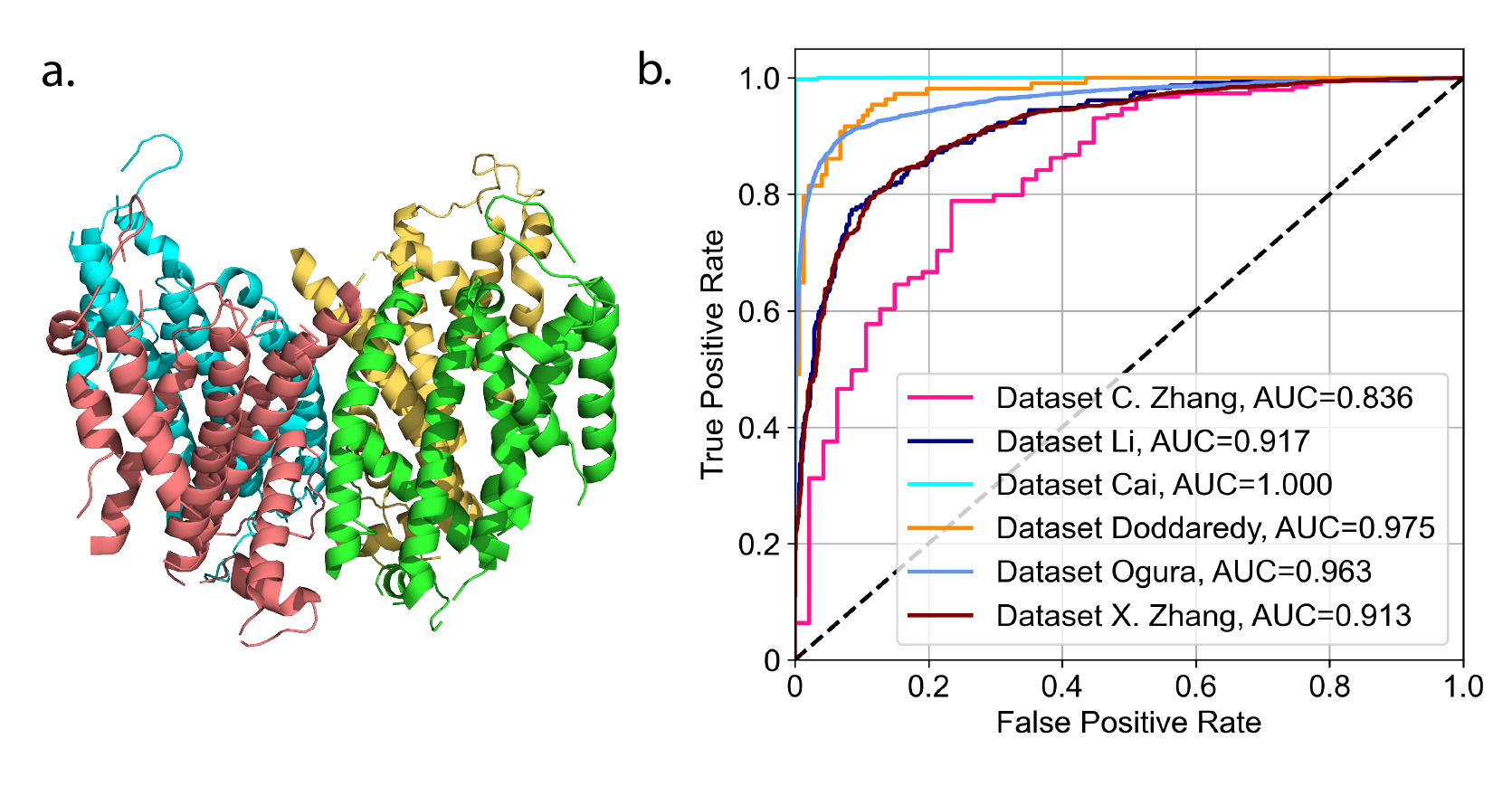} 
	\caption{{\footnotesize a. The Cryo-EM structure of hERG protein with PDB code 7CN1. It has four identical $\alpha$-subunits with each containing six $\alpha$-helical transmembrane domains. b. The ROC-AUC curve of our classification models for six datasets with separate test sets. } }
	\label{Fig:AUC-6datasets}
\end{figure}

Braga et al. \cite{braga2015pred} collected a curated hERG dataset composed of 3388 blockers and 3436 non-blockers.  Based on the dataset, they developed their Pred-hERG models, which showed an accuracy of 0.78 in the 5-fold cross validation task. In X. Zhang et al.'s  \cite{zhang2022hergspred} recently proposed work, the accuracy was improved to 0.814 in the 5-fold cross-validation task. Based on the same training set, our consensus model achieved the same accuracy for 5-fold cross validation predictions. In terms of other evaluation metrics as shown in Table \ref{tab:classication-results}, our consensus model delivered comparable or superior results compared with earlier models. The AUC, MCC, and accuracy results are close or identical, whereas our specificity result was higher than that of \cite{zhang2022hergspred}. C. Zhang et al. \cite{zhang2016silico} investigated their model performance with a small hERG dataset containing 1163 compounds. Five models were developed and their prediction performances were compared on different training and test sets partitioned from the 1163 compounds. Four thresholds defined by IC$_{50}$ values, namely 1 $\mu$M, 5 $\mu$M, 10 $\mu$M, and 30 $\mu$M  were considered to discriminate hERG blockers from non-blockers. Their SVM model had the best predictive ability for the test set determined by the threshold of 30 $\mu$M with a reported accuracy of 0.848 on this test set. With the same training and test set, X. Zhang's model \cite{zhang2022hergspred} had prediction improvement with accuracy boosted to 0.856. Our consensus model had much higher improvement in terms of almost every metric on the same test set. Our model enhanced the accuracy from 0.856 to 0.864 while our AUC and MCC results have large increases from 0.803 to 0.836 and from 0.445 to 0.518, respectively. Li et al.  \cite{li2017modeling} constructed two consensus models based on their dataset composed of 3721 compounds with a threshold of IC$_{50}$ equals to 1 $\mu$M classifying blockers and non-blockers. Their best consensus results on a test set of 1092 compounds achieved an accuracy of 0.842. Our consensus model provided a significant improvement with respect to most metrics compared to the results of Li et al. \cite{li2017modeling} and X. Zhang et al.\cite{zhang2022hergspred}. Our AUC, accuracy and, MCC results are 0.917, 0.885 and, 0.629, which are much higher than the best result of 0.881, 0.867, and 0.584 from Li et al. \cite{li2017modeling} or X. Zhang et al. \cite{zhang2022hergspred}on the three same metrics. In the work of Cai et al.   \cite{cai2019deep}, a multitask deep neural network-based (MT-DNN) model was proposed and showed outstanding performance compared to a series of machine learning or deep learning models. Their MT-DNN model demonstrated the best predictive power on the test set with the threshold value of 80 $\mu$M. The reported AUC and accuracy results achieved 0.967 and 0.925. With the same training and test sets, our consensus model achieved nearly perfect evaluation metrics values equaling to or close to 1.0. In particular, the MCC values were raised from 0.92 to 0.99. Doddaredy et al.'s \cite{doddareddy2010prospective} hERG dataset is a popular benchmark dataset and was used to examine the performance of many models in predicting hERG blockade classification. Wang et al.'s model \cite{wang2020capsule} proposed two deep neural network-based models, namely Conv-CapsNet and RBM-CapsNet, which achieved the best performance on Doddaredy et al.'s test set compared to all other literature models. The training and test set of the Doddaredy et al.'s hERG dataset consists of 2389 and 255 compounds, respectively. In the training and test sets, compounds with IC$_{50}$ less than 10$\rm \mu$M are classified as blockers while IC$_{50}$ value greater than 30$\rm \mu$M defines compounds without activity. By comparison, our consensus model had a boosted AUC value of 0.975, which is much higher than the reported AUC value of 0.944 in \cite{wang2020capsule}. Ogura et al. \cite{ogura2019support} built their support vector machine (SVM) model with the largest hERG blockade dataset by integration of compounds from ChEMBL, GOSTAR, PubChem, and the NIHChemical Genomics Center (NCGC). The training and test consist of 203,853 and 87,366 compounds, respectively. IC$_{50}$ of 10$\mu$M was used as the blocker/non-blocker threshold. Ogura et al.'s model \cite{ogura2019support} gave accuracy and AUC of 0.984 and 0.962. Our consensus model had comparable accuracy or slightly better AUC results compared to Ogura et al.'s model. Another large hERG blockade dataset was provided in Ref. \cite{zhang2022hergspred} by integrating literature dataset and the data in the ChEMBL database.  X. Zhang et al.'s \cite{zhang2022hergspred} model had a reported accuracy of 0.839, while our prediction had a higher accuracy of 0.842. 
Through comprehensive comparisons with other models on several datasets, our consensus model consistently demonstrated excellent predictions and is among the best machine learning models of the hERG blocker/non-blocker classifier.

AUC is an important evaluation metric in classification studies. It has the advantage of measuring the model's prediction quality regardless of the discrimination threshold. The ROC curves of our model's prediction for the six datasets with separate test sets are shown in Fig. \ref{Fig:AUC-6datasets}.

\subsection{Molecular embedding analysis }

\begin{figure}[ht]
	\centering
	\includegraphics[width=0.9\linewidth]{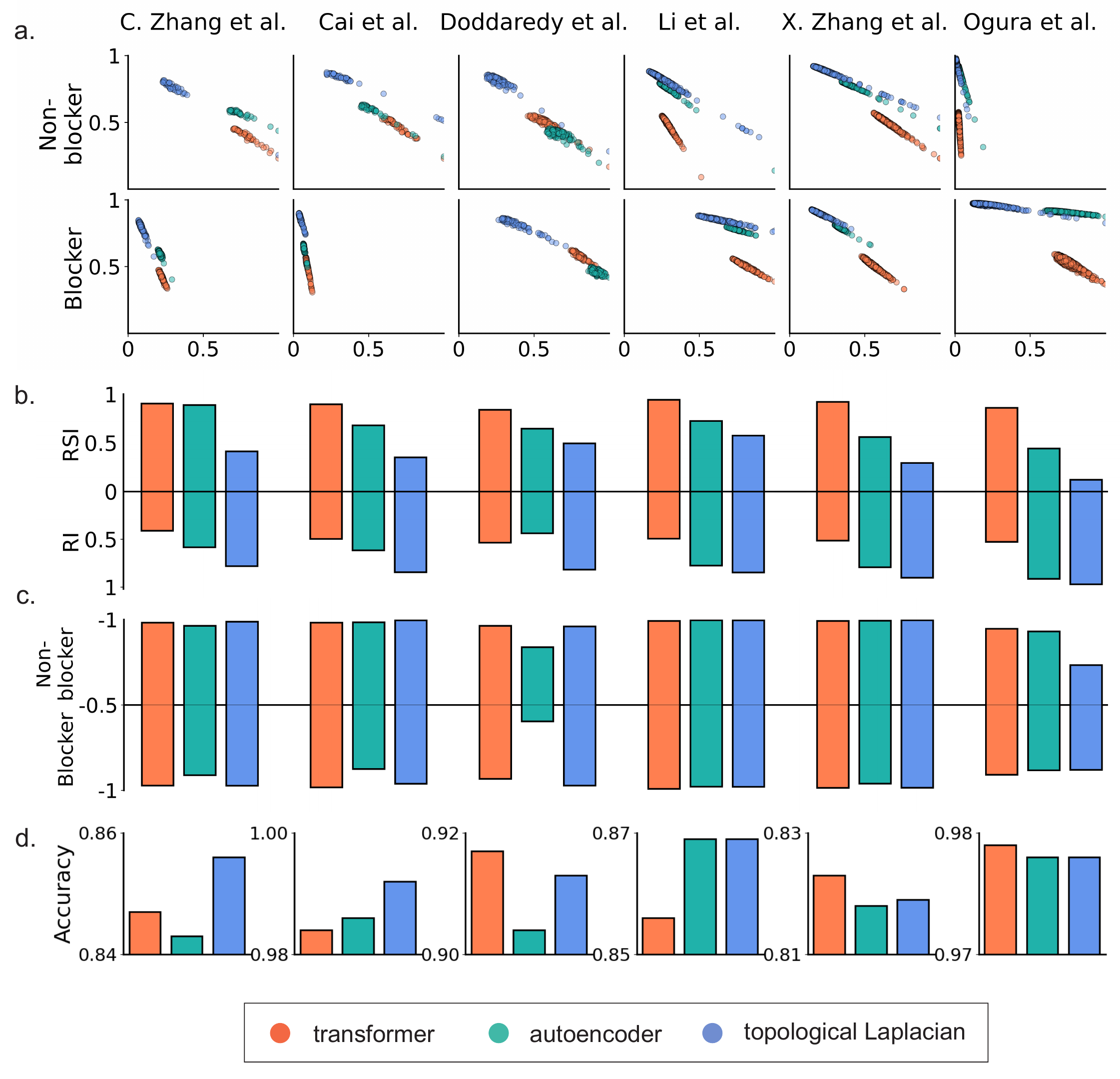} 
	\caption{  Illustration of the R-S score, SI, RSI, RSC, and accuracy of transformer (TF), autoencoder (AE), and topological Laplacian (TL) embeddings for six data sets. 
		a. R-S score analysis of TF, AE, and TL.  
		b. SI and RSI analysis of TF, AE, and TL. 
		c. RSC analysis of TF, AE, and TL. 
		d. Accuracy  of TF, AE, and TL.  	
	}  
	\label{Fig:rs-score}
\end{figure}

The residue index (RI), similarity index (SI) and R-S index (RSI) are three indexes derived from residue-similarity scores \cite{hozumi2022ccp}. We would like to compare the individual performance of the three molecular embeddings in the binary classification tasks. Fig. \ref{Fig:rs-score}a shows the plots of residue-similarity scores for the three types of embeddings. The $x$-and $y$-coordinates stands for the residue-similarity scores, respectively. We have six datasets with separate test sets. The residue-similarity scores of molecules in one test set were calculated by comparing to the molecules in corresponding training sets. Such comparisons involves distances or similarity calculations in terms of embeddings within or across classes. The formulation of aforementioned scores or indexes can be found in the method section. In our previous investigation \cite{hozumi2022ccp}, the RSI was speculated to have high correlation with the prediction accuracy in classification tasks. The classification study for hERG blocker/non-blocker is a binary problem. The Fig. \ref{Fig:rs-score}d records the accuracy comparisons of the three embeddings for the six classification tasks. Transformer and topological Laplacian embeddings achieved the best accuracy for three and four tasks, respectively. They had very close performance in Ogura et al.'s \cite{ogura2019support} dataset with prediction accuracy difference of 0.001. It is shown in Fig. \ref{Fig:rs-score}b that the transformer embedding achieved the highest RSI values across all the six datasets. On the other hand, TL embedding has the highest similarity index values for all the six datasets. This suggests that both RSI and SI can be crucial indicators of high prediction accuracy in such binary classification tasks. Fig. \ref{Fig:rs-score}c shows the correlation between residue and similarity scores (RSC) of the three embeddings. Their correlation coefficients are all negative for the three embeddings in the prediction of  all the blocker/non-blocker classes. The magnitudes of RSC are all high enough. Especially, those by transformer embedding are close to 1 for all datasets. Beside, the RSC by autoencoder and topological Laplacian embeddings are close to -1 for most datasets.

\section{Materials and Methods}

\subsection{Datasets}

In this study, seven hERG blockade datasets were used to evaluate the predictive performance of our classification models. Based on the same datasets, we compared the performance of our models with other published models. The seven datasets include Braga et al.'s  \cite{braga2015pred},  C. Zhang et al.'s \cite{zhang2016silico}, Li et al.'s \cite{li2017modeling}, Cai's \cite{cai2019deep}, Doddaredy \cite{doddareddy2010prospective}, X. Zhang et al.'s  \cite{zhang2022hergspred}, Ogura et al.'s \cite{ogura2019support}. The oldest one was collected by Doddaredy et al.'s in 2010 while the most recently collected one was from X. Zhang \cite{zhang2022hergspred} in 2022. Five out of the seven datasets have less than 10,000 sample points in their training set whereas the other two have training set of more than 10,000 compounds. It is noted that the one by Ogura et al.'s \cite{ogura2019support} is the largest hERG blockade dataset to date. Categories of hERG blockers and non-blockers are defined differently in terms of various $\rm IC_{50}$ thresholds in the seven datasets. A frequently used threshold is to classify compounds with $\rm IC_{50}\leq$ 10$\mu$M  as blockers and those with $\rm IC_{50}>10\mu$M as non-blockers or decoys. Different thresholds separate distinct pools of blockers and decoys, which affect discrimination prediction by machine learning. The prediction performance of our models was assessed with other published models based on the given split training and test set. The statistics of the seven datasets are shown in Table \ref{Tab:dataset-info}. The partitions for training and test sets with the corresponding number of blocker/non-blocker compounds are specified. 

We used the seven datasets to evaluate the performance of our machine learning models in classifying hERG blockers/non-blockers. The three types of embeddings were used to build machine learning models for the seven datasets except for Ogura et al.'s dataset \cite{ogura2019support}. Additional molecular embedding from algebraic graph learning (AGL) \cite{nguyen2019agl} was employed to build models for Ogura et al.'s dataset \cite{ogura2019support} and to further enhanced predictive performance. The consensus probability results from eight models determined the classification predictions of hERG blockers/non-blockers. A brief description of AGL is provided in the Supporting information. The purpose of this study is to screen DrugBank compounds on their hERG-blocking potential. We need one hERG inhibitor dataset that covers compounds of enough chemical and structural diversities. This consideration follows the fact that the hERG channel can be blocked by drugs from diverse chemical and therapeutic classes. The dataset by Ogura et al. \cite{ogura2019support} is our top choice. We utilized the classification model based on this dataset to screen DrugBank database for potential hERG blockers. In addition to the qualitative analysis with the classification model, we built regression models to predict the binding affinity of these potential blockers. We collected an hERG inhibitor dataset of 6298 compounds from the ChEMBL database. With this dataset, our regression model was built. 

\begin{table}[ht]
	\centering
	\small
	\begin{tabular}{c c c c c }		
		\toprule
		Dataset & Splitting& Blockers  &   Non-blockers&   Total\\ 
		\hline
		\multirow{2}*{Braga \cite{braga2015pred}} & 5-fold&	\multirow{2}*{3388}& \multirow{2}*{3436} &\multirow{2}*{6824}\\
		&cross validation & &   &		 \\
		\hline
		
		\multirow{2}*{C. Zhang \cite{zhang2016silico}}  & Training &	562 & 365 &	927\\
		&Test & 244&  163 &407\\
		\hline
		
		\multirow{2}*{Li \cite{li2017modeling}} & Training &	973& 2748 &	3721	\\
		&Test & 234&  858 &1092  \\
		\hline
		
		\multirow{2}*{Cai \cite{cai2019deep}} & Training &	3485& 469 &	3954	\\
		&Test & 435&  58 &493 \\
		\hline
		
		\multirow{2}*{Doddaredy\cite{doddareddy2010prospective}} & Training &	1004& 1385 &	2389	\\
		&Test & 108&  147 &255 \\
		\hline
		
		\multirow{2}*{Ogura \cite{ogura2019support}} & Training &	6923& 196930 &	203853	\\
		&Test & 2967&  84399 &87366\\
		\hline
		
		\multirow{2}*{X. Zhang \cite{zhang2022hergspred}} & Training &	5530& 4750 &	10859	\\
		&Test & 1377& 1193 &2570  \\
		\bottomrule
	\end{tabular}
	\caption{Details of seven hERG datasets used for benchmarking our classification models. The Braga et al.'s  dataset\cite{braga2015pred} is used for five-fold cross validation while all others have separate test sets.}
	\label{Tab:dataset-info}
\end{table}

\subsection{Molecular embeddings}

In this study, three types of molecular embeddings were mainly deployed to build machine learning models. Each embedding was generated based on different philosophy. Their collective performance through consensus results demonstrated extraordinary predictive power in the aforementioned classification tasks. Besides, their individual performances were compared in the last section. In the following, we give a comprehensive description of the three types of embeddings.

\paragraph{Topological Laplacians  }

Graph theory investigates the relationship between objects including nodes, edges, faces and high-dimensional generalizations. The explorations are being made with a range of tools from geometric graph theory, algebraic graph theory, topological graph theory, and spectral graph theory. Among them, spectral graph theory unveils both topological invariants and homotopic shape information through the harmonic and non-harmonic spectra of the Laplacian matrix, respectively. Another aspect of spectra analysis is by the de Rham-Hodge theory that is based on differential geometry. The Hodge Laplacians for Laplace-Beltrami operator on a compact Riemannian manifold allows us to obtain topological and geometric insight of the underlying manifold by harmonic and non-harmonic spectra. The evolutionary de Rham-Hodge theory provides more detailed analysis of evolving manifolds defined through filtration parameters \cite{chen2021evolutionary}. Filtration parameter is a tool to generate a series of geometric shapes for a given data, on which, for example, persistent homology extracts the topological persistence as one popular topological data analysis (TDA) \cite{zomorodian2004computing,xia2014persistent}. The evolutionary de Rham-Hodge theory develops the analysis with differential geometry and algebraic topology while persistent homology fulfills the analysis via multiscale analysis and algebraic topology. 

Similar to the evolutionary de Rham Hodge theory, topological Laplacians (TLs), including persistent spectral graphs \cite{wang2020persistent}, persistent path Laplacians \cite{wang2022persistent}, and persistent sheaf Laplacian \cite{wei2021persistent},  form families of persistent $q$-combinatorial Laplacian operators. The harmonic and non-harmonic spectra can be obtained from these Laplacians to capture the topological invariants and homotopic shape evolution of the data, respectively.   Topological Laplacians are powerful tools for the multiscale analysis of the topological invariants and homotopic  shape evolution of data.

A filtration of an oriented simplicial complex $K$ is a sequence of the sub-complexes $\{K_t\}_{t=0}^m$ of $K$:
\begin{align*}
	\emptyset =K_0 \subseteq K_1 \subseteq K_2 \subseteq \cdots \subseteq K_m =K.
\end{align*}
On each simplicial complex $K_t$, the chain complex is defined as $C_q^t:=C_q(K_t)$ and the $q$-boundary operator $\partial_q^t: C_q(K_t)\rightarrow C_{q-1}(K_t)$ exists. For the general case with $0<q\leq$ dim$(K_t)$, the $q$-boundary operator takes the form
\begin{align*}
	\partial_q^t(\sigma_q)=\sum\limits_i^q(-1)^i\sigma_{q-1}^i,\enspace\text{for}\enspace \sigma_q\in K_t,
\end{align*}
where $\sigma_q=[v_0,v_1,\cdots,v_q]$ is an oriented $q$-complex, and $\sigma_{q-1}^i=[v_0,\cdots,\hat{v}_i,\cdots,v_q]$ is an oriented $(q-1)$-simplex with the vertex $v_i$ removed. When $q<0$, the $C_q(K_t)=\{\emptyset\}$ and $\partial_q^t$ is a zero map. Corresponding the $q$-boundary operator, an adjoint operator called $q$-adjoint boundary operator  is defined as
\begin{align*}
	\partial_q^*:C_{q-1}(K_t)\rightarrow C_{q}(K_t).
\end{align*}
Consider $\mathbb{C}_q^{t+p}$, a subset of $C_q^{t+p}$ with its boundary in $C_{q-1}^t$:
\begin{align*}
	\mathbb{C}_q^{t+p}:=\{\sigma \in C_q^{t+p}| \partial_q^{t+p}(\sigma)\in C_{q-1}^t\}.
\end{align*}
On this subset, the $p$-persistent $q$-boundary operator $\crpartial_q^{t+p}:\mathbb{C}_q^{t+p}\rightarrow C_{q-1}^t$ and the adjoint boundary operator $(\crpartial_q^{t+p})^*:C_{q-1}^{t}\rightarrow \mathbb{C}_{q}^{t+p}$ are well defined. The $p$-persistent $q$-combinatorial Laplacian operator
\begin{align*}
	\Delta_q^{t+p}=\crpartial_{q+1}^{t+p}\left(\crpartial_{q+1}^{t+p}\right)^*+(\partial_q^t)^*\partial_q^t,
\end{align*}
which has following matrix representations:
\begin{align*}
	\mathcal{L}_q^{t+p}=\mathcal{B}_{q+1}^{t+p}\left(\mathcal{B}_{q+1}^{t+p}\right)^T+\left(\mathcal{B}_q^t\right)^T\mathcal{B}_q^t.
\end{align*}
It is noted that matrices $\mathcal{B}_{q+1}^{t+p}$ and $\mathcal{B}_{q}^{t}$ are the matrix representations for boundary operators $\crpartial_{q+1}^{t+p}$ and $\crpartial_{q}^{t}$, respectively. The number of rows in $\mathcal{B}_{q+1}^{t+p}$ equals that of oriented $q$-simplices in $K_t$, and the column number corresponds to the number of oriented $(q+1)$-simplices in $K_{t+q}\cap \mathbb{C}_{q+1}^{t+p}$. Besides, the transposes of $\mathcal{B}_{q+1}^{t+p}$ and $\mathcal{B}_{q}^{t}$ are the matrix representation for $\left(\crpartial_{q+1}^{t+p}\right)^*$ and $(\partial_q^t)^*$. It is known that the topological and spectral information of $K_t$ can be revealed from the Laplacian operator. We obtain the spectra of $\mathcal{L}_q^{t,p}$ and denote the set of spectra as
\begin{align*}
	\text{Spectra}\left(\mathcal{L}_q^{t+p}\right) = \left\{(\lambda_1)_q^{t+p}, (\lambda_2)_q^{t+p},\cdots, (\lambda_N)_q^{t+p}\right\},
\end{align*}
where $N$ indicates the dimension of $\mathcal{L}_q^{t+p}$. 
The Betti numbers, the number of zero eigenvalues, of Laplacian matrix can reveal $q$-cycle information. For the $p$-persistent $q$-combinatorial Laplacian matrix $\mathcal{L}_q^{t+p}$, the Betti number is defined as below:
\begin{align*}`
	\beta_q^{t+p}=\text{dim}\left(\mathcal{L}_q^{t+p}\right)-\text{rank}\left(\mathcal{L}_q^{t+p}\right)=\text{nullity}\left(\mathcal{L}_q^{t+p}\right)=\text{number of zero eigenvalues of }\mathcal{L}_q^{t+p}.
\end{align*}
The $\beta_q^{t+p}$ value record the number of $q$-cycles in $K_t$ that are still alive in $K_{t+p}$. For the 3D structure problem, the order of $q$ ranges from 0 up to 2 as 0, 1, and 2 indicates vertex, edges, and faces, respectively. Correspondingly, $\beta_q^{t+p}$ value measures the persistence of connected components, tunnels or circles, and cavities or voids. The harmonic persistent spectra can be used to track the topological changes while non-harmonic persistent spectra enable us to derive geometric changes. This made TL even more powerful than persistent homology that solely deals with the topological invariants.

\begin{figure}[ht!]
	\centering
	\includegraphics[width=0.9\linewidth]{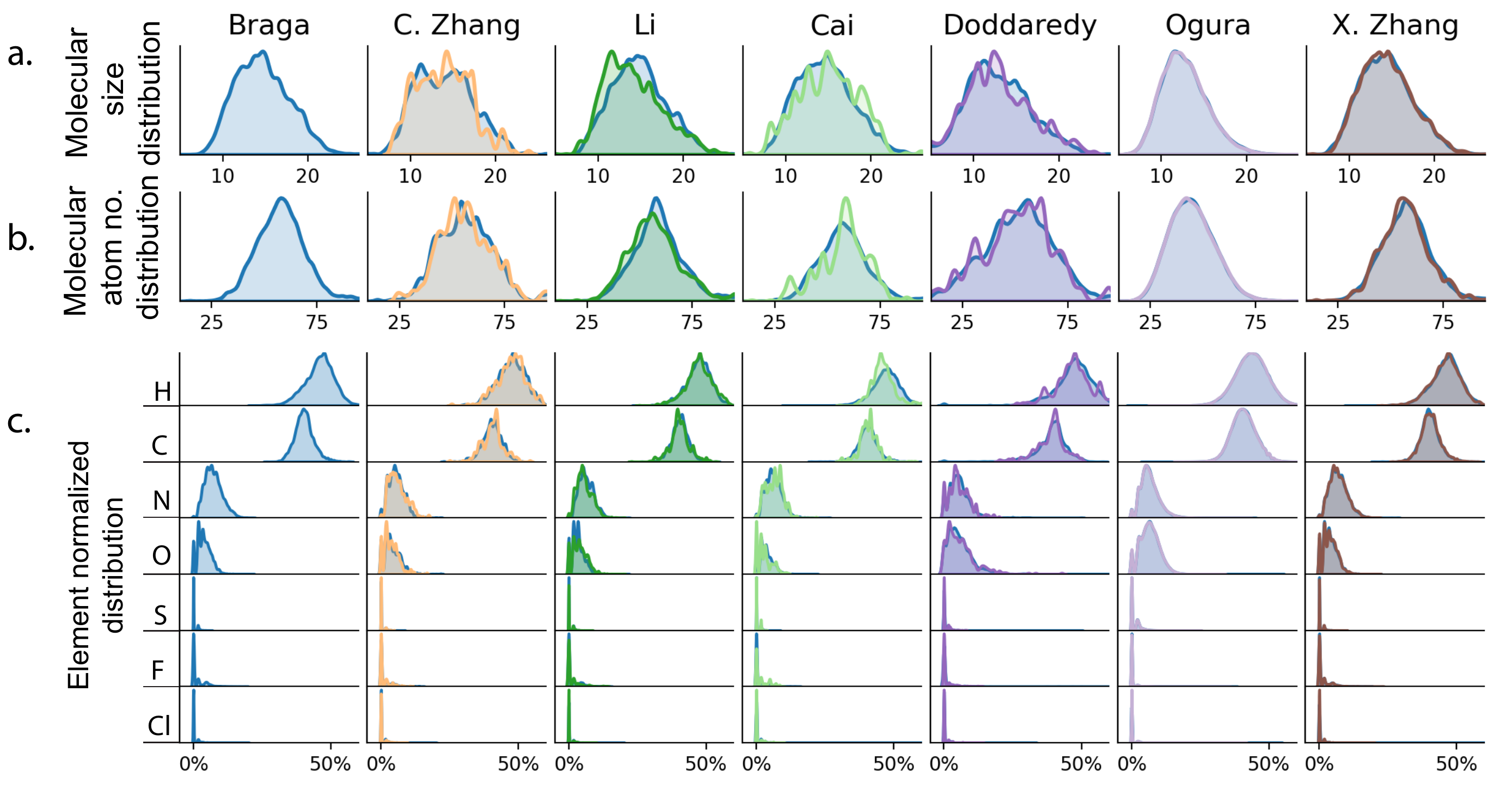} 
	\caption{{\footnotesize The distributions analysis of the seven classification datasets. The blue color indicates the training sets of each dataset while other colors denote the distributions for corresponding test sets. a. shows the molecular size distribution, where size is measured by the maximum molecular length in the three Cartesian directions. b. represents atom number distributions of molecules for the seven datasets. c. represents the distributions of element portion in single molecule for the seven datasets. The most occurring atoms in molecules come from the listed seven types of elements. Atoms of element H and C are the two dominating constituents in molecules while the portion of other kinds of atoms are low.} }
	\label{Fig:distribution}
\end{figure}

In the framework of TL, statistics on eigenvalue from Laplacian matrix $\mathcal{L}_0^{t+p}$ is used to generate molecular features. The statistics include $\beta_0^{t+p}$ and the sum, mean, median, maximum, minimum, standard deviation, variance, sum of the square of the non-harmonic spectra. The performance with TL feature relies on the selection of atoms from element combinations, which constitutes different oriented $q$-simplices in $K_t.$ Element-specific Laplacian matrices were constructed with different combinations of atoms along the filtration radius. It requires us to analyze the element types and their atomic proportion in molecular dataset. Besides, in this study, the filtration radius takes lower and upper bound of 1 and 10 angstrom, respectively. This is elucidated by the molecular size distribution as shown in Fig. \ref{Fig:distribution}a. Most compounds in each dataset have 3D size of less than 20 angstroms in the Cartesian directions. More details of element-specific Laplacian matrices based on distribution analysis can be found in the Supporting information.

\paragraph{Sequence-to-sequence auto-encoder  }

Recently, Winter et al \cite{winter2019learning} proposed a data-driven unsupervised learning  model  to extract molecular information embedded in the SMILES representation. Sequence-to-sequence autoencoders was utilized to translate one form of molecular representation to another form, with comprehensive description of chemical structure compressed in latent representation between encoder and decoder. The resulting model allows for the molecular descriptors extraction for query compound without retraining or using labels. 

The translation model consists of the encoder and decoder networks. Both convolutional neural network (CNN) and recurrent neural network (RNN) architectures were tried and then fully connected layers maps the output of CNN or the concatenated cell states of the RNN networks to intermediate vector representations between encoders and decoders. The decoder contains RNN networks with the latent representations as input. In order to compress more molecular chemical properties in the latent vectors, an additional classification model was built by mapping the latent vectors to molecular property vectors. The mean squared errors were defined to measures the molecular property predictions. The decoder's RNN network gave the probability distributions over different characters for the translated molecular sequences. The loss function for training the autoencoder model is defined to be the sum of cross-entropy between probability distributions and hot-hot encoded correct characters and the mean squared errors for molecular property predictions. The translation model was trained with approximately 72 million molecular compounds from ZINC and PubChem databases. Meaning molecular descriptors were formed in the latent vectors when the translation model achieved high accuracy. We call such molecular descriptors in latent vector as auto-encoder (AE) embedding in our study.

\paragraph{Bidirectional Transformer  }

Our previous work \cite{chen2021extracting} developed a self-supervised learning (SSL)-based platform that extracts molecular descriptors by bidirectional encoder transformer (BET). Massive molecular SMILES from three databases including ChEMBL, PubChem, and Zinc were used as input in the transformer-based SSL pretraining, leading to learned representations of molecules. Following the idea of the BERT model \cite{devlin2018bert} for natural language processing, our BET model only used the encoder architecture. But the input to the encoder was molecular SMILES string and only the masked learning task was kept in the pretraining process. The transformer encoder has the advantage of high parallelism and is particularly beneficial for training massive data.

To fulfill the SSL, preprocessing of SMILES was needed. A total of 51 symbols were considered as components of the SMILES string. In particular, symbols $'\langle s \rangle'$ and  $'\langle \backslash s \rangle'$ were added to the beginning and end of the SMILES as input to the encoder. The maximum length of SMILES is 256. The symbol $'\langle pad \rangle'$ is used to supplement a SMILES string if its length is less than 256. A total of 15$\%$ of the symbols were operated, among which $80\%$ were masked, $10\%$ were unchanged and the remaining $10\%$ were randomly changed. The BET framework recovered the symbols of the masked parts by learning the unprocessed parts of SMILES in the training process. The learned representation for molecular SMILES string can be achieved. 

The BET consists of eight bidirectional encoder layers, with each encoder containing a multi-head self-attention layer and a fully connected feed-forward neural network. The self-attention layers play a key role in the transformer models and capture the importance of symbols. The attention mechanism used in each encoder layer is scaled dot-product attention as described follows
\begin{align}\label{seq:self-attention}
	\text{Attention}\left(Q, K,V\right)=\text{Softmax}\left(\frac{QK^T}{\sqrt{d_k}}\right)V,
\end{align}
where $Q$, $K$ and $V$ are the query matrix, key matrix, and values matrix. The $d_k$ stands for the embedding dimension of each token for the symbols The scaling factor $\sqrt{d_k}$ applied on the dot product of query and key matrix counteracts the divergence of dot-product. The softmax function 
In our study, the multi-head self-attention is composed of 8 self-attention headers and each header draws attention on different aspects of the symbol embedding and is beneficial to the prediction performance. The residual connection and layer normalization techniques were applied to each of the two sub-layers. The cross-entropy function defined the loss function that measured the difference between the predicted and real symbols at masked positions. The Adam optimizer was used in the training process with the weight decay set to 0.1. The embedding dimension of each symbol was 512 and the embedding size of fully connected feed-forward layers was 1024. The molecular embedding matrix is composed of 256 embedding vectors with dimension 512. The mean of embedding vectors for the valid symbols in one SMILES string is formed as our molecular descriptors, which can be used in downstream machine learning tasks. It is hence taken as our bidirectional transformer embedding. Three pretrained models were generated with molecular SMILES from one or the union of the ChEMBL, PubChem, and ZINC databases. In this study, the pretrained model solely using the ChEMBL database was employed to generate transformer-based embeddings.

\subsection{Residue-Similarity (R-S) scores and indexes}

Residue-Similarity (R-S) scores and indexes was proposed in our previous work \cite{hozumi2022ccp}  and is a potentially powerful visualization tool in data science. It can serve as a an alternative visualization approach in addition to traditional dimensionality reductions including principal component analysis (PCA) and uniform manifold approximation and projection (UMAP), but it has the advantage of avoiding aggressive dimension reduction that induces poor representations for high-dimension data. Its visualization applications for classification problems  is not limited to binary study.

Residue-Similarity (R-S) scores or R-S plots are composed of two components, residue and similarity scores. Assume the interested data samples forms a set $\Omega=\{(x_m,y_m)|x_m \in \mathbb{R}^N, y_m \in \mathbb{Z}_L\}_{m=1}^M$, where $x_m$ is the $m$th data point. The label $y_m$ indicates the ground truth in classification problem while it represents the cluster label in clustering problem. The feature representation $x_m \in \mathbb{R}^N$ has $N$ features and the dataset has $M$ data samples. Besides, $L$ means that the data have $L$ types of labels, namely, $y_m \in [0,1,2,\cdots,L]$ . The whole set $\Omega$ can be partitioned into $L$ classes 
$\omega_l=\{x_m \in \Omega|y_m=l\}$ according to the labels $y_m=l$ and hence $\Omega= \cup_{i=0}^{L-1} w_l.$

The residue score is defined to be the normalized inter-class sum of the distances. Suppose $x_m \in \omega_l,$ then the inter-class sum of the distances is given as
\begin{align*}
	R(x_m) = \sum\limits_{x_j\notin \omega_l} ||x_m-x_j||,
\end{align*}
where $||\cdot||$ defines the distance metric for a pair of vectors. Then the residue score for $x_m$ is equal to 
\begin{align}
	R_m := \frac{1}{R_{\max}}R(x_m),
\end{align}
where $R_{\max}=\max \limits_{x_m \in \Omega}R(x_m).$ The similarity score $S_m$ is the average of the intra-class scores. Specifically, for any $x_m \in \omega_l$,
\begin{align}
	S_m := \frac{1}{|\omega_l|} \sum\limits_{x_j\in \omega_l} \left(1-\frac{||x_m-x_j||}{d_{\max}}\right),
\end{align}
where $d_{\max}= \max \limits_{x_i,x_j\in\Omega}||x_i-x_j||$. It is noted that both the residue score and similarity score range between 0 and 1. In our study, the Euclidean distance is mainly used for the R-S scores, while others can be employed as well. Generally, a large residue score $R_m$ indicates the data has large dissimilarity from data in other classes, while a large similarity score $S_m$ indicates that the data is well-clustered in the same class. 

The residue score and similarity score can be employed in the visualization of each class, where $\{R_m\}_{m=1}^{M}$ and $\{S_m\}_{m=1}^{M}$ are the x- and y-coordinates respectively. To compare the overall performance in the clustering or classification prediction using different feature representations $\{x_m\}_{m=1}^{M}$, we define the class residue index (CRI) and class similarity index for each class. For a class $\omega_l$, the two indexes are defined as CRI$_l= \frac{1}{|w_l|}\sum_m R_m$ and CSI$_l= \frac{1}{|C_l|}\sum_m S_m.$ with their range of $[0,1]$. Considering the $L$ classes for the data, it is more useful to define residue index (RI) and similarity index (SI) for the whole set as RI$=\frac{1}{L}\sum_l $CRI$_l$ and SI$=\frac{1}{L}\sum_l $CSI$_l$. It is intuitively true that it is better if these indexes are large no matter for the class-independent ones or the overall indexes. We can also define the R-S disparity (RSD) with RSD=RS-SI and furthermore we define R-S index (RSI) with RSI=1-$\rm |RI-SI|$. All these indexes have a range of $[0,1]$ except that RSD $\in[-1,1]$. We also define the R-S correlation (RSC) between residual $\{R_m\}_{m=1}^{M}$ and similarity $\{S_m\}_{m=1}^{M}$ scores  with the formula
\begin{align}
	\text{RSC}= \frac{\sum_{m=1}^{M}(R_m-\bar{R})(S_m-\bar{S})}{\sqrt{\sum_{m=1}^{M}(R_m-\bar{R})^2}\sqrt{\sum_{m=1}^{M}(S_m-\bar{S})^2}},
\end{align}
where $\bar{R}$ and $\bar{S}$ are the mean of $\{R_m\}_{m=1}^{M}$ and $\{S_m\}_{m=1}^{M}$, respectively. As residue and similarity are two opposite measures of data analysis, the RSC has range of $\in[-1,0]$.

\subsection{Model construction}
In the last subsection, advanced mathematics-based,  self-supervised and unsupervised learning-based molecular embeddings have been presented. Those embeddings were combined with machine learning algorithms to develop our models. Gradient boosting decision tree and deep neural network (DNN) algorithms were adopted in our study. The generated three types of embeddings combined with two machine learning algorithms give rise to totally six individual models, from which we obtain consensus models for hERG blockade predictions. The consensus results are formed by the simple average of the probability score. It has been found that the consensus approach can boost the machine learning prediction in a variety of molecular property studies \cite{gao2021proteome,li2017modeling}, and typically outperforms each individual models. 
\begin{table}
	\centering
	\small
	\begin{tabular}{c |c| c |c}		
		\toprule
		Dataset &  GBDT parameters & DNN hyperparameters  & DNN batch size\\ 
		\hline
		Braga \cite{braga2015pred} & $\rm n{\_}estimators = 30000$  &	 & 64\\
		
		C. Zhang \cite{zhang2016silico} &$\rm max{\_}depth = 7$ &  &8	\\
		
		Li \cite{li2017modeling} & $\rm min{\_}samples{\_}split = 5$  &	 $L_2$ normalization penalty: 0.0005&16\\
		
		Cai \cite{cai2019deep} &$ \rm subsample  = 0.8$  & 	Momentum: 0.9& 16\\
		
		Doddaredy\cite{doddareddy2010prospective} &$\rm learning{\_}rate = 0.002$ &Activation: Relu, softmax  &	16\\
		
		X. Zhang \cite{zhang2022hergspred} & $\rm max{\_}features = "sqrt"$  & Learning rate: 0.001  & 8	\\
		\cline{1-2}
		\multirow{6}*{Ogura \cite{ogura2019support}} & $\rm n{\_}estimators = 100000$  &  Optimizer: SGD& \\
		
		&$\rm max{\_}depth = 7$ &Iteration epoch: 200  &\\
		
		& $\rm min{\_}samples{\_}split = 3$  &	&	\\
		
		&$ \rm subsample  = 0.3$  & &16 \\
		
		&$\rm learning{\_}rate = 0.01$ &  &	\\
		
		& $\rm max{\_}features = "sqrt"$  &	&\\
		
		\bottomrule
	\end{tabular}
	\caption{Hyperparameter setting of GBDT and DNN for all the seven classification datasets.}
	\label{tab: GBDT-DNN-params}
\end{table}

\paragraph{Gradient boosting decision tree}

Gradient boosting decision tree (GBDT) is an ensemble algorithm widely used for regression and classification tasks. Its philosophy lies in creating a large number of weak learners (individual trees) by bootstrapping training samples and making predictions by integrating the outputs of weak learners.   Weak learners are likely to make mistakes in predictions. The ensemble approach allows for reducing the overall error by combining all the weaker learners. The GBDT algorithm is less sensitive to hyperparameters, less prone to overfitting, and easy to implement. When trained on small datasets, it can deliver better performance than DNN and a variety of other machine learning algorithms. It gained popularity in a wide range of quantitative structure-activity relationship
prediction problems \cite{wu2018quantitative, cang2018integration}. The GBDT packages provided in scikit-learn (version 0.24.1) library were used in this work. Two sets of GBDT hyperparameters were employed with the details summarized in Table \ref{tab: GBDT-DNN-params}.

\paragraph{Deep neural network}

The neural network typically consists of multiple interconnected layers of neutrons and mimics the human brain to solve problems with numerous neuron units with backpropagation to update weights on each layer. The input layer has a neuron number equaling to the length of the input vector encoded by the molecular feature representations. The hidden neuron layers record weighted sum of the output from its previous layer. Deep learning abstracts more properties of the molecular features through network layers and the neuron in each layer. The last neuron layer yields the prediction of the models. Our binary classification only contains one neuron for the prediction output. In this study, all DNN models are composed of three hidden layers with varying input layer sizes determined by the number of input features. The first and second hidden layers had double size of the input layer size. The third layer had the same size as the input layer.  The stochastic gradient descent (SGD) algorithm is used for optimization with the $\rm momentum$ parameter set as 0.9. The $L_2$ normalization was used for regularization to avoid overfitting. The penalty weight was set to be 0.0005. The network is fully connected with no dropout used.  Due to the different sizes of datasets, no uniform batch size was used. The rectified linear unit (ReLU) was used as the activation function between the neuron layers except for the last layer. The softmax activation was used for the last layer. $L_2$-penalized binary cross entropy as described in Eq. (\ref{eq:BCE}) was used as the loss function
\begin{align}\label{eq:BCE}
	{\rm   BCE} = -\frac{1}{N}\sum_{i=1}^{N}[y_i\cdot \log(p(y_i)) + (1-y_i)\cdot \log(1-p(y_i))]+\lambda ||W||_2^2
\end{align}
where $p(y_i)$ is the probability of class blocker, $1-p(y_i)$ is the probability of class non-blocker, $N$ indicates the number of compounds in the training set,  $||\cdot||_2$ represents the $L_2$ norm, and $\lambda$ denotes the penalty weight. The epoch numbers for all datasets are set to 200. The DNN hyperparameters for all datasets are also listed in Table \ref{tab: GBDT-DNN-params}. All the DNN models were implemented with PyTorch (version 1.10.0).

\section{Conclusion}

In this work, we construct new machine learning models to screen the DrugBank database for potential hERG blockers. A few sets of advanced molecular embeddings were integrated with gradient boosting tree and deep neural network algorithms to build  state-of-art machine learning models. Two sequence-based molecular embeddings were generated by two natural language processing (NPL) methods while two 3D structure-based molecular embeddings were devised by advanced mathematics including topological Laplacians and algebraic graphs. The sequence-based embeddings are complementary to the 3D structure-based ones in representing molecules. Their collective efforts of describing molecules give rise to exceptional predictive ability of machine learning models. Our models shed light on the side effect of DrugBank compounds on the hERG channel.  According to our classification models, 227 DrugBank compounds were predicted to be potential hERG blockers. Our regression models give the binding affinity predictions for these blockers. Some of the 227 predicted blockers have reported cardiotoxicity or hERG-blocking induced side effects, confirming our predictions. Our predictions offer a timely warning of the potential side effects of these blockers. Further experimental tests or clinical trials are urgently needed to scrutinize the alerted hERG-liability and cardiotoxicity.

\section*{Data  and code availability}

The related datasets studied in this work are available at: 
\url{https://weilab.math.msu.edu/DataLibrary/2D/}. Codes are available at \url{https://github.com/WeilabMSU/hERG-prediction}.

\section*{Acknowledgment}
This work was supported in part by NIH grants  R01GM126189 and R01AI164266, NSF grants DMS-2052983,  DMS-1761320, and IIS-1900473,  NASA grant 80NSSC21M0023,  MSU Foundation,  Bristol-Myers Squibb 65109, and Pfizer.

\end{document}